\documentclass[14pt, twoside, a4paper]{article} %                      
\usepackage{amsmath}

\usepackage{tikz}
\usepackage{graphicx}
\usepackage{subfigure}
\usepackage{calc}
\usepackage{float}

\let\getprepared\relax%            %%
\ifx\TestIngCommand\undefined\relax%                                       %%
\else\input ../../../proc/ppreamb-book.tex %                               %%
  \input init.tex \fi%                                                     %%
\let\TestIngCommand\undefined%                                             %%

\newtheorem{remark}{Remark}

\usepackage{amsmath,amscd,amssymb}                                         %%
\usepackage{graphics}                                                     %%
\newtheorem{theo}{Theorem}                                                 %%
\newtheorem{lem}{Lemma}                                                    %%
\newtheorem{defi}{Definition}
\newskip\ttglue\ttglue=.5em plus.25em minus.15em                           %%
\def\firstname#1{\def\FIRSTNAME{#1}\ignorespaces}%                         %%
\def\lastname#1{\def\LASTNAME{#1}\ignorespaces}%                           %%
\def\middleinitial#1{\def\MIDDLEINI{#1}\ignorespaces}%                     %%
\def\department#1{\def\DEPARTMENT{#1}\ignorespaces}%                       %%
\def\institute#1{\def\INSTITUTE{#1}\ignorespaces}%                         %%
\def\address#1{\def\ADDRESS{#1}\ignorespaces}%                             %%
\def\country#1{\def\COUNTRY{#1}\ignorespaces}%                             %%
\def\otheraffiliation#1{\def\OTHERAFFILIATION{#1}\ignorespaces}%           %%
\def\email#1{\def\EMAIL{#1}\ignorespaces}%                                 %%
\newcount\autcount\autcount=0                                              %%
\newcount\affcount\affcount=0                                              %%
\newcount\numcount\newcount\nummcount                                      %%
\newcount\nummmcount\newcount\nummmmcount                                  %%
\def\writename#1#2{\ \kern-1ex\hbox{%                                      %%
  \csname AUthor\the#1\endcsname\                                          %%
  \edef\TESTSTR{}\expandafter\ifx\csname auTHor\the#1\endcsname\TESTSTR    %%
  \else\csname auTHor\the#1\endcsname.\ \fi                                %%
  \csname authOR\the#1\endcsname$^{\csname AFF\the#1\endcsname}$%          %%
  \expandafter\ifx\csname corr\number#1\endcsname\relax                    %%
  \else\thanks{Corresponding author.}\ \fi                                 %%
  }\ifnum#1<#2, \else\ \kern-1ex\fi}%                                      %%
\def\writeemail#1{%                                                        %%
  \nummcount=0\relax\nummmcount=0\relax                                    %%
  \loop\ifnum\nummcount<\autcount\advance\nummcount by1\relax              %%
    {\expandafter\ifnum\csname AFF\the\nummcount\endcsname=#1\relax        %%
    \global\advance\nummmcount by1\fi}\repeat                              %%
  \nummcount=0\relax\nummmmcount=0\relax                                   %%
  \loop\ifnum\nummcount<\autcount\advance\nummcount by1\relax              %%
    {\expandafter\ifnum\csname AFF\the\nummcount\endcsname=#1\relax        %%
    \global\advance\nummmmcount by1\relax\def\blank{}\expandafter          %%
    \ifx\csname EMAIL\the\nummcount\endcsname\blank(no e-mail)%            %%
    \else\csname EMAIL\the\nummcount\endcsname                             %%
    \fi                                                                    %%
    \ifnum\nummmmcount<\nummmcount; \fi\fi}\repeat}%                       %%
\long\def\BeginAuthorList#1\EndAuthorList{#1\relax                         %%
  \author{\vbox{\hsize=390pt\noindent\numcount=0\relax                     %%
    \loop\ifnum\numcount<\autcount\advance\numcount by1\relax              %%
      \writename{\numcount}{\autcount}%                                    %%
      \repeat}\\[2mm]                                                      %%
    \vbox{\small\numcount=0\relax                                          %%
      \loop\ifnum\numcount<\affcount\advance\numcount by1\relax            %%
        \vbox{{\count0=\numcount\relax                                     %%
          \loop\expandafter\ifnum\csname AFF\the\count0\endcsname%         %%
            <\numcount\relax\advance\count0 by1\relax\repeat               %%
          $^{\csname AFF\the\count0\endcsname}$}%                          %%
        \def\BLANK{}\expandafter\ifx\csname DEPT\the\numcount\endcsname    %%
          \BLANK                                                           %%
          \else\csname DEPT\the\numcount\endcsname, \fi                    %%
        \csname INST\the\numcount\endcsname,                               %%
        \csname ADDR\the\numcount\endcsname,                               %%
        \csname COUN\the\numcount\endcsname                                %%
        \edef\TEST{}\expandafter\ifx\csname OTHE\the\numcount\endcsname%   %%
          \TEST                                                            %%
          .\else;\break\csname OTHE\the\numcount\endcsname.\fi}%           %%
        \vbox{\writeemail{\numcount}}%                                     %%
        \repeat}\\}}%                                                      %%
\expandafter\def\csname x1\endcsname{}%                                    %%
\expandafter\def\csname x2\endcsname{}%                                    %%
\expandafter\def\csname x3\endcsname{}%                                    %%
\expandafter\def\csname x4\endcsname{}%                                    %%
\expandafter\def\csname x5\endcsname{}%                                    %%
\expandafter\def\csname x6\endcsname{}%                                    %%
\expandafter\def\csname x7\endcsname{}%                                    %%
\expandafter\def\csname x8\endcsname{}%                                    %%
\expandafter\def\csname x9\endcsname{}%                                    %%
\def\Author#1#2{\global\advance\autcount by1\relax#2                       %%
  \expandafter\edef\csname AUthor\the\autcount\endcsname{\FIRSTNAME}%      %%
  \expandafter\edef\csname auTHor\the\autcount\endcsname{\MIDDLEINI}%      %%
  \expandafter\edef\csname authOR\the\autcount\endcsname{\LASTNAME}%       %%
  \expandafter\edef\csname EMAIL\the\autcount\endcsname{\EMAIL}%           %%
  \let\tempera\"\def\"{\string\"}\expandafter\ifx\csname x\DEPARTMENT%     %%
    \endcsname\relax                                                       %%
    \global\advance\affcount by1\relax\let\"\tempera                       %%
    \expandafter\edef\csname DEPT\the\affcount\endcsname{\DEPARTMENT}%     %%
    \expandafter\edef\csname INST\the\affcount\endcsname{\INSTITUTE}%      %%
    \expandafter\edef\csname ADDR\the\affcount\endcsname{\ADDRESS}%        %%
    \expandafter\edef\csname COUN\the\affcount\endcsname{\COUNTRY}%        %%
    \expandafter\edef\csname OTHE\the\affcount\endcsname{\OTHERAFFILIATION}%%
    \expandafter\edef\csname AFF\the\autcount\endcsname{\the\affcount}%    %%
  \else\expandafter\edef\csname AFF\the\autcount\endcsname{\DEPARTMENT}%   %%
  \fi\let\"\tempera\ignorespaces}%                                         %%
\def\CorrespondingAuthor#1#2{%                                             %%
  \expandafter\xdef\csname corr\number#1\endcsname{cor}%                   %%
  \Author#1{#2}}%                                                          %%
\def\PaperTitle#1{\title{\bf#1}}%                                          %%
\def\Category#1{\ignorespaces}%                                            %%
\def\keywords#1{{\noindent \emph{Keywords:}                                %%
  \def\BLANK{}\def\TEST{#1}\ifx\BLANK\TEST(n/a).\else#1\fi}}%              %%
\getprepared                                                               %%
\begin{document}                                                           %%

\PaperTitle{A Mathematical Analysis of Benford's Law and its Generalization}
\Category{(Pure) Mathematics}

\date{}

\BeginAuthorList
 \Author1{
    \firstname{Alex}
    \lastname{Kossovsky}
    \middleinitial{E}   %% One letter, no period (.); or leave empty.
    \department{Author and Inventor}
    \institute{Independent Consultant}
    \otheraffiliation{}
    \address{New York}
    \country{USA}
    \email{akossovsky@gmail.com}}
  \Author2{
    \firstname{Wayne}
    \lastname{Lawton}
    \middleinitial{M}   %% One letter, no period (.); or leave empty.
    \department{Department of the Theory of Functions, Institute of Mathematics and Computer Science}
    \institute{Siberian Federal University}
    \otheraffiliation{}
    \address{Krasnoyarsk}
    \country{Russian Federation}
    \email{wlawton50@gmail.com}}
\EndAuthorList
\maketitle
\thispagestyle{empty}
\begin{abstract}
We explain Kossovsky's generalization of Benford's law which is a formula that approximates the distribution of leftmost digits in finite sequences of natural data and apply it to six sequences of data including populations of US cities and towns and times between earthquakes.
We model the natural logarithms of these two data sequences as samples of random variables having normal and reflected Gumbel densities respectively. We show that compliance with the general law depends on how nearly constant the periodized density functions are and that the models are generally more compliant than the natural data. This surprising result suggests that the generalized law might be used to improve density estimation which is the basis of statistical pattern recognition, machine learning and data science.
\end{abstract}
%
% 42A16 Fourier coefficients, Fourier series of functions with special properties
% 60E05 Distributions: generaltheory
% 62-04 Explicit machine computation and programs
% 62G07 Densityestimation
%
\footnote{Kossovsky authored mathematical studies and books on Benford's Law, invented an algorithm (patent US9058285B2), and consulted on data forensics.}
\footnote{\thanks{This work of the second author Lawton 
is supported by the Krasnoyarsk Mathematical Center and financed by 
the Ministry of Science and Higher Education of the Russian Federation (Agreement No. 075-02-2023-936).}}
\section{Introduction}\label{sec1}
In 1938 Frank Benford proposed (\cite{ben}, Eqn 1) that the first or leftmost digits  $d = 1,...,9$ in natural data are not uniformly distributed, instead their frequencies are approximately
$$\log_{10}(1+1/d) \approx  \{ 0.301, 0.176, 0.125, 0.097, 0.079, 0.067, 0.058, 0.051, 0.046 \}.$$
This distribution of digits is called Benford's Law. It was earlier discovered by Simon Newcomb in 1881 \cite{newcomb} who noticed that in logarithm tables the earlier pages (that started with 1) were much more worn than the other pages.
\\ \\
In (\cite{koss4}, Section 8) Kossovsky
demonstrated that Benford's approximation is accurate for finite sequences of positive-valued natural data with sufficient variability. 
For $\delta \in (0,1)$ and for a finite sequence of positive numbers
$s$ define $T_\delta(s)$ to be the sequence $s$ truncated by removing the fraction $\delta$ of smallest terms and fraction $\delta$ of largest terms, and define the ratio
\begin{equation}\label{Rdelta}
	R_\delta(s) := \frac{\hbox{max }(T_\delta(s))}{\hbox{min }(T_{\delta}(s))}
\end{equation}
In (\cite{koss4}, Section 9) he proposed $R_{0.01}(s)$ as a measure of variability of a sequence and observed that Benford's approximation is very accurated when $\log_{10} R_{0.01}(s) \geq 3.$
We will discuss the robustness of this statistic in our empirical analysis in Sections 4 and 5. 
\\ \\
Benford's law extends to base-$(D+1) \geq 3$ to give probabilities 
$\log_{D+1}(1+1/d), d = 1,...,D$ that depend on partitioning positive numbers into D subsets determined by the first digit in their base-(D +1) postional notation representation. Very accurate Benford approximation is expected when $\hbox{length}(s)  >> D$ and 
\begin{equation}\label{outstandingBenford}
	\log_{D+1} R_{0.01}(s) \geq 3. 
\end{equation}
In this paper iff means if an only if, $:=$ means that the expression to its left is defined to be the quantity to its right. $\mathbb Z, \mathbb Q, \mathbb R, \mathbb C, \mathbb R_+$ are the integer, rational, real, complex and positive numbers.
$\ln : \mathbb R_+ \mapsto \mathbb R$ is the natural logarithm function.
\section{The General Law}\label{sec2}
In 2013 Kossovsky generalized Benford's partition \cite{koss1} and in more detail in (\cite{koss2}, Section 7) by creating a partition described in Appendix 1 where we use it to derive, for $F > 1$ and integer $D \geq 2,$ the partition of $\mathbb R_+$ below:
\begin{equation}\label{SigmaFD}
\Sigma_{F,D}(d) := \bigcup_{j \in \mathbb Z}  [1+(d-1)(F-1)/D,1+d(F-1)/D)F^j, \ \  d = 1,...,D.
\end{equation}
He defined the rank function
$R_{F,D}: \mathbb R_+ \mapsto \{1,...,D\}$ by
\begin{equation}\label{Rank}
 R_{F,D}(x) := d \hbox{ such that } x \in \Sigma_{F,D}(d), \ \  x \in \mathbb R_+, \, d = 1,...,D. 
\end{equation}
If $F = D+1$ then $R_{F,D}(x)$ is the leftmost digit in the base-$F$ representation of $x.$ For every finite $\mathbb R_+$-valued sequence 
$s = (s_1,...,s_n)$ Kossovsky defined rank frequencies
\begin{equation}\label{rankfreq}
	Freq(R_{F,D}(s) = d) := 
	\frac{1}{n}\, card \, \{j : R_{F,D}(s_j) = d\}, \ \  d = 1,...,D
\end{equation}
and observed that for data sequences with length $n \geq 2000$ 
having sufficient variability, and using $F$ and $D$ values roughly between 2 and 17, the rank frequencies are
\begin{equation}\label{GLapprox}
	Freq(R_{F,D}(s) = d) \approx L_{F,D}(d)
\end{equation}
where
\begin{equation}\label{LFD}
L_{F,D}(d) := \log_F \, \frac{1+d(F-1)/D}{1+(d-1)(F-1)/D}, \ \ d = 1,...,D
\end{equation}
and he named approximation (\ref{GLapprox}) the General Law Of Relative Quantities (GLORQ) (\cite{koss2}, p. 561). For the sake of brevity we will refer to this as the General Law (GL). Very close GL approximation is expected when $\hbox{length}(s)  >> D$ and 
\begin{equation}\label{outstandingGL}
	\log_F R_{0.01}(s) \geq 3. 
\end{equation}
In \cite{koss2} he quantified the error in the GL approximation (\ref{GLapprox}) by the sum of squares deviation (SSD) statistic
\begin{equation}\label{Devdisc}
	E_{F,D}(s) := 
\sum_{d=1}^D \left[ \, Freq(R_{F,D}(s) = d) - L_{F,D}(d) \, \right]^2
\end{equation}
\begin{defi}\label{flat}
A finite sequence $s$ is $(F,D)$-flat if 
$Freq(R_{F,D}(s) = d) = \frac{1}{D}, \ d = 1,...,D.$ 
\end{defi}
\begin{lem}\label{antiGL1}
If $s$ is $(F,D)$-flat then
\begin{equation}\label{antiGL2}
	D \times E_{F,D}(s) := D \sum_{d = 1}^D (L_{F,D}(d))^2 - 1.
\end{equation}
\end{lem}
Proof: Follows since $\Sigma_{d=1}^D L_{F,D}(d) = 1.$
\\ \\
The intermediate value theorem gives
\begin{equation}\label{Lbounds}
	\frac{(F-1)/ln F}{1+d(F-1)/D} \leq D \times L_{F,D}(d) < \frac{(F-1)/ln F}{1+(d-1)(F-1)/D}, \ d = 1,...,D
\end{equation}
which implies that $L_{F,D}(d)$ is a strictly decreasing function of $d$ and
\begin{equation}\label{limD1}
	\lim_{D \rightarrow \infty} \frac{L_{F,D}(D)}{L_{F,D}(1)}  = \frac{1}{F}.
\end{equation}
Define
\begin{equation}\label{SF}
	S(F) := \frac{(F-1)^2}{F(\ln F)^2}.
\end{equation}
An integral comparison gives:
\begin{equation}\label{limD2}
	\bigg| \, D \sum_{d = 1}^D (L_{F,D}(d))^2 -  S(F) \, \bigg| 
\leq \frac{1}{D}\frac{F-1}{F\ln F}
\end{equation}
The table below shows values of $S(F)-1$ for selected values of $F.$
$$
\begin{array}{cccccccccc}
F        & 1.01    & 1.1       & 1.5      &   2       & 8 & 32 & 128 & 512 \\
S(F)-1 & 8.3e-6 &	 7.6e-04 & 1.4e-2 &  4.1e-2 & 4.2e-1 & 1.5 & 4.4 & 12
\end{array}
$$
\begin{remark}\label{flattest}
Applying L'H\^{o}pital's rule gives that
$\lim_{F \rightarrow 1} L_{F,D}(d) = \frac{1}{D}, \ , d = 1,...,D.$ Therefore
$\lim_{F \rightarrow 1} S(F)-1 = 0,$ as shown in the table.
If a sequence $s$ is flat then 
$D \times E_{F,D}(s) \approx S(F)-1.$ Therefore the table together with the preceding results provides a means of testing the null hypothesis that a sequence is flat.
\end{remark}
\begin{lem}\label{lem1}
If $D_1$ divides $D_2$ then $D_1\times E_{F,D_1}(s) \leq D_2\times E_{F,D_2}(s)$ for every finite positive-valued sequence $s.$
\end{lem}
Proof: Let $m := D_2/D_1$ and $F_j(d_j) := Freq(R_{F,D_j}(s) = d_j), j = 1,2.$ Then
$$F_1(d_1) - L_{F,D+1}(d_1) = 
\sum_{d_2 = m(d_1-1)+1}^{md_1}  [F_2(d_2) - L_{F,D_2}(d_2)], \ d_1 = 1,...,D_1$$
The Schwarz inequality (\cite{rud1}, Theorem 11.35) gives
$$\bigg|F_1(d_1) - L_{F,D+1}(d_1)\bigg|^2 \leq
m \sum_{d_2 = m(d_1-1)+1}^{md_1}  [F_2(d_2) - L_{F,D_2}(d_2)]^2.$$
The result follows by summing over $d_1$ as follows:
$$D_1 \times E_{F,D_1}(s) = D_1 \sum_{d_1 = 1}^{D_1} \bigg|F_1(d_1) - L_{F,D+1}(d_1)\bigg|^2$$
$$\leq D_1m \sum_{d_1 = 1}^{D_1}\sum_{d_2 = m(d_1-1)+1}^{md_1}  [F_2(d_2) - L_{F,D_2}(d_2)]^2$$
$$ = D_2 \sum_{d_2 = 1}^{D_2} [F_2(d_2) - L_{F,D_2}(d_2)]^2 = D_2 \times  E_{F,D_2}(s).$$
\section{General Law for Natural Data}\label{sec3}
The three tables below show the quantities
$L_{F,D}(d), F = 2, 8, 32; d = 1,...,5;$ and for 
six  sequences $s$ of natural data the quantities
$Freq(R_{F,5}(s) = d), F = 2, 8, 32; d = 1,...,5.$ The last columns show $E_{F,D}(s).$
The seven lines in each of the three tables below refer to:
\begin{enumerate}
\item $L_{F,5}(d), F = 2, 8, 32; d = 1,..5.$
\item Masses of $n = 1404$ known exoplanets in the Milky Way Galaxy as of September 21, 2016, with the measurements of mass in units of Jupiter. $\sigma(\ln s) = 2.26,$ $R_{0.01}(s) = 15482.$ 
\item Market capitalization values of the list of $n = 2889$ companies registered on the NASDAQ Exchange as of the end of Oct 9, 2016. 
$\sigma(\ln s) = 2.09,$ $R_{0.01}(s) = 18001.$ 
\item Populations of $n = 19509$ incorporated cities and towns reported in the US 2009 census. 
$mean(\ln s) = 7.2170,$ $\sigma(\ln s) = 1.8316,$ $R_{0.01}(s) = 3678.$ 
\item Prices in US dollars of $n = 15194$ electronics items in an online catalogue of Canford PLC in October 2013.
$\sigma(\ln s) = 2.09,$ 
$R_{0.01}(\widetilde s) = 9306.$
\item House numbers in $n = 23633$ street addresses in Prince Edward Island, Canada.  
$\sigma(\ln s) = 2.13,$ 
$R_{0.01}(s) = 12621.$ 
\item Time in seconds between $n = 19451$ successive earthquakes worldwide for the year of $2012.$ 
$mean(\ln s) = 7.2170,$ $\sigma(\ln s) = 1.3754,$ $R_{0.01}(s) = 629.$
\end{enumerate}
$$
\begin{array}{ccccccc}
  L_{2,5}    & 0.263 & 0.222 & 0.193 & 0.170   & 0.152 & 0\\
  Planets     & 0.262 &	0.226 & 0.188 & 0.171   & 0.152 & 0.00004\\
  Capital.     & 0.264 &	0.216 & 0.195 & 0.169   & 0.155 & 0.00005\\
  US Pop.     & 0.262 &	0.223 & 0.193 & 0.176   & 0.147 & 0.00006 \\
  Prices       & 0.257 &	0.225 &	0.197 & 0.161 & 0.160 & 0.00021 \\
  Addresses & 0.263 &	0.217 &	0.197 & 0.165 & 0.158	 & 0.00010 \\
Earthquakes & 0.268 &  0.222  & 0.192  & 0.167  & 0.152 & 0.00003
\end{array}
$$
$$
\begin{array}{ccccccc}
  L_{8,5}    & 0.421 & 0.221 & 0.151 & 0.115 & 0.093  & 0 \\
  Planets     & 0.419 &	0.204	& 0.170 & 0.108 & 0.100  & 0.00075 \\
  Capital.     & 0.414 &	0.230	& 0.155 & 0.115 & 0.087  &  0.00018 \\
  US Pop     & 0.423 & 0.221	& 0.148 & 0.114 & 0.093  & 0.00001 \\
  Prices       &  0.427 & 0.205 & 0.154 & 0.124 & 0.090 & 0.00038 \\
  Addresses &  0.417 & 0.226 & 0.159 & 0.114 & 0.084 & 0.00017 \\ 
Earthquakes & 0.411  & 0.225  & 0.160  & 0.113  & 0.091 & 0.00021
\end{array}
$$
$$
\begin{array}{ccccccc}
  L_{32,5}   & 0.570 & 0.179 & 0.110  & 0.079  & 0.062 &  0 \\
  Planets     & 0.593 & 0.223	& 0.108   & 0.046  & 0.031 &  0.00461 \\
  Capital.     & 0.564 & 0.186  & 0.115  & 0.073  & 0.062  & 0.00014 \\
  US Pop     & 0.600  & 0.168	 & 0.102	 & 0.070	& 0.060  & 0.00121
 \\
  Prices       &  0.589 & 0.161 & 0.114  & 0.077   & 0.060  & 0.00076 \\
  Addresses & 0.606  & 0.168  & 0.095	 & 0.078	&  0.053  & 0.00177 \\
Earthquakes & 0.509 &  0.221  & 0.131  & 0.086  & 0.053 & 0.00592
\end{array}
$$
\begin{remark}
In (\cite{koss3}, p. 142, 159) Kossovsky computed the values in the first table for $F = 2.$ We substituted earthquake data for his data that consisted of molar masses in grams of $n = 2175$ commonly used and naturally occurring chemical compounds for two reasons. First, the GL error was significantly more for molar data than other data. Second, there is a combinatorial chemistry reason why molar data does not follow the GL. Simple molecules such a $H_2$ are more numerous than complex molecules such as DNA, but there are far fewer types of simple molecules. For example the hydrocarbon propane $C_3H_8,$ with molar mass $44.097g,$ has one type. The hydrocarbon butane $C_4H_{10},$ with molar mass $58.12,$ has two isomers, one with a straight carbon chain and the other with one carbon bonded with three carbons. This is why the molar mass data was not included in the above tables for this article.
\end{remark}
The three tables show two striking features:
\begin{enumerate}
\item The frequencies $Freq(R_{F,5}(s) = d), d = 1,...,5$ decrease with increasing $d$ and are very close to the theoretical $L_{F,D}(d), d = 1,...,5 $ values.
\item The quantities $E_{F,D}(s)$ are small but increase significantly as $F$ increases.
\end{enumerate}
\section{Population Model}\label{sec4}
Figure 1 displays a histogram of the natural logarithm of the population data. It appears to be normal so we modeled the data as random samples of a random variable $X$ that has a lognormal distribution and $\ln X$ has density
\begin{equation}\label{normalp}
	p(x) = \frac{1}{\sigma \sqrt {2\pi}} 
\exp\left[-\frac{1}{2}\left(\frac{x-\mu}{\sigma} \right)^2\right]
\end{equation}
where $\mu = 7.2170$ and $\sigma = 1.8316$ are the mean and standard deviation of the natural logarithms of the data sequence.
To test our model we compared the cumulative distribution of $\ln s$
with the theoretic cumulative distribution of $\ln X$ in Figure 2 and computed the Kolmogorov-Smirnov statistic
\begin{equation}\label{KS1}
KS := \sqrt n \times \max |\hbox{ difference between cumulative distributions }|
\end{equation}
to have value 6.1457. Kolmogorov's theorem implies that as $n$ increases $KS$ converges to the random variable $\sup_{t \in [0,1]} |B(t)|$ where $B(t)$ is the Brownian bridge, thus
\begin{equation}\label{KS2}
\lim_{n \rightarrow \infty}
\hbox{Prob } (KS \geq x) = 2\sum_{k=1}^\infty (-1)^{k-1} e^{-2k^2x^2}
\end{equation}
Since our value $n = 19509$ is rather large we obtain
\begin{equation}\label{KS3}
\hbox{Prob } (KS \geq x) \approx 2\sum_{k=1}^\infty (-1)^{k-1} e^{-2k^2x^2}
\end{equation}
hence 
\begin{equation}\label{KS4}
 \hbox{Prob } (KS \geq 6.1457) \approx 1.5621 \times 10^{-33}
\end{equation}
suggesting we should decisively reject the null hypothesis that the data was formed from independent random samples of a lognormal distribution with the specfified parameters. 
\\ \\
Moreover, for our lognormal distribution, 
$R_{0.01}(X) = \exp(-2 \times 2.3263 \times \sigma) = 5022$ 
compared to our empirical value $R_{0.01}(s) = 3678.$ The reason these values differ is because the samples are not chosen from an exactly lognormal density as proved by the Kolmogorov-Smirnov test. Nevertheless the lognormal approximation is rather accurate.
\begin{figure}
    \centering
    \includegraphics[width=4in]{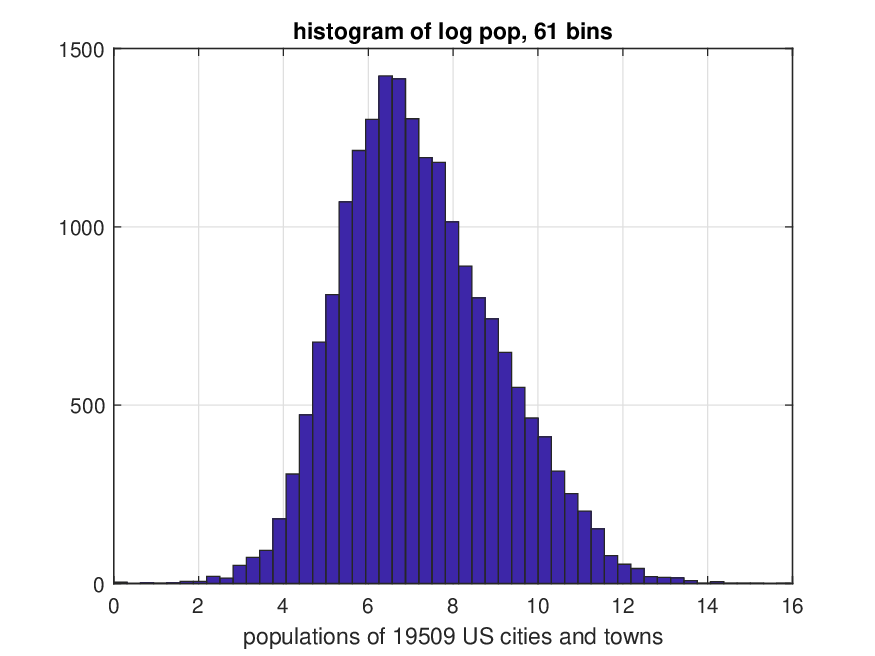}
    \caption{Populations Histogram}
    \label{pop}
\end{figure}
\begin{figure}
    \centering
    \includegraphics[width=4in]{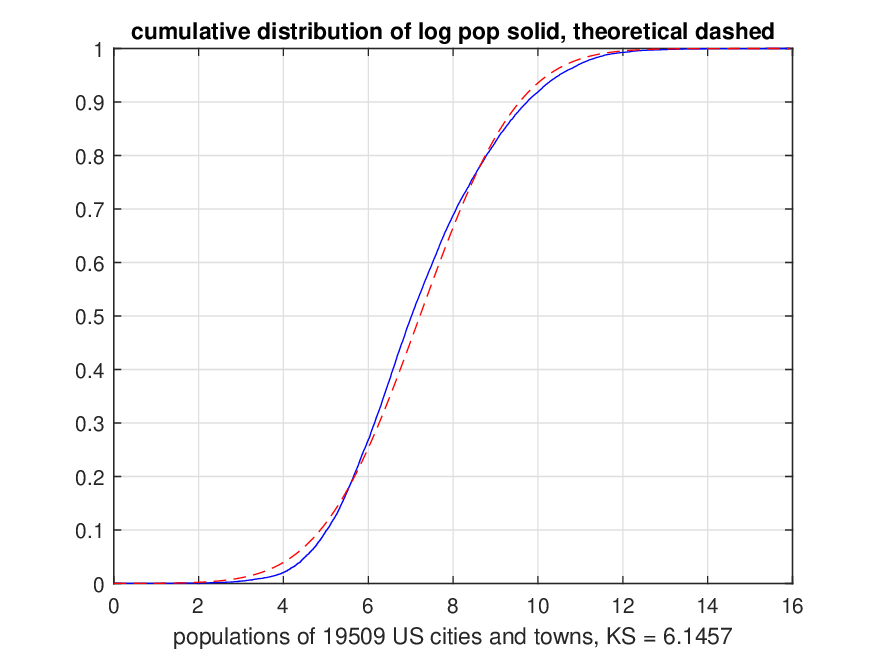}
    \caption{Populations Cumulative}
    \label{pop}
\end{figure}
\section{Earthquake Time Model}
Figure 3 displays a histogram of the natural logarithm of times between earthquake data. It appears be reflected Gumbel distributed so we modeled the data as random samples of a random variable $X$ such that $\ln X$ has density
\begin{equation}\label{refGumbel}
	p(x) = \frac{1}{\beta} \exp (z - e^z), \ z = \frac{x + \mu}{\beta}
\end{equation}
where $\mu \in \mathbb R$ and $\beta >0.$ Its standard deviation equals $\pi\beta/\sqrt 6$ and its mean equals 
$\mu - \beta \gamma$ where $\gamma \approx 0.5772$ is the 
Euler-Mascheroni constant. Using our data mean $7.2170$ and standard deviation $1.3754$ for $\ln s$ gives
$\beta = 1.0724$ and $\mu = 7.8360.$
To test our model we compared the cumulative distribution of $\ln s$
with the theoretic cumulative distribution of $\ln X$ in Figure 4 and computed the Kolmogorov-Smirnov statistic
to have value $1.8058.$ Thus
\begin{equation}\label{KS5}
 \hbox{Prob } (KS \geq 1.8058) \approx 0.0015
\end{equation}
which suggests rejecting the null hypothesis that the data was formed from independent random samples of a log reflected Gumbel distribution with the specified parameters. However, the rejection is far less compelling than for the population data. The distributions of earthquake times is far closer to the log reflected Gumbel distribution than the distribution of populations is to the lognormal distribution.
\\ \\
Moreover, for our log reflected Gumbel
$R_{0.01}(X) = 714$ 
compared to our empirical value $R_{0.01}(s) = 629$ which is much closer than for the population data.
\begin{figure}
    \centering
    \includegraphics[width=4in]{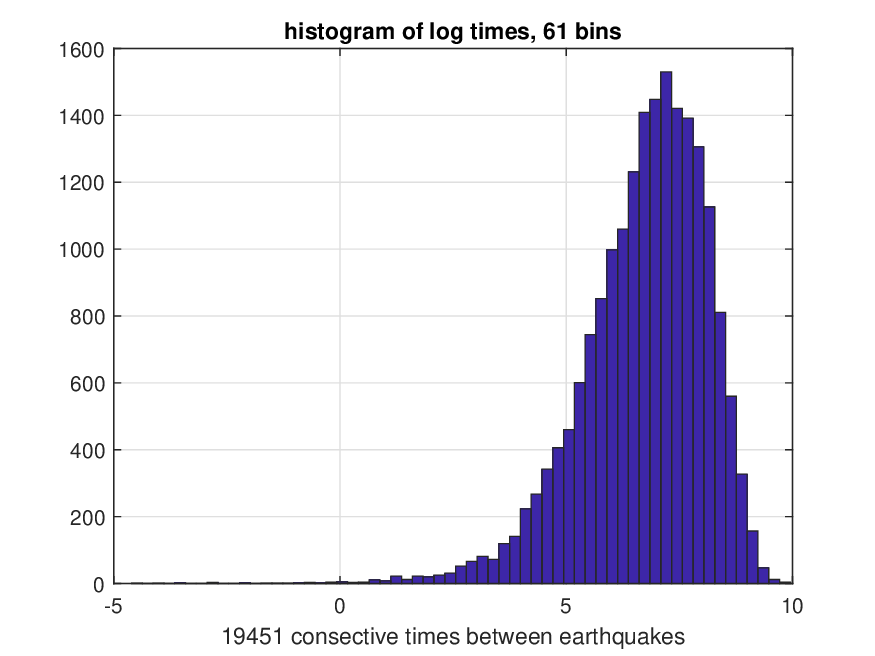}
    \caption{Earthquake Times Histogram}
    \label{pop}
\end{figure}
\begin{figure}
    \centering
    \includegraphics[width=4in]{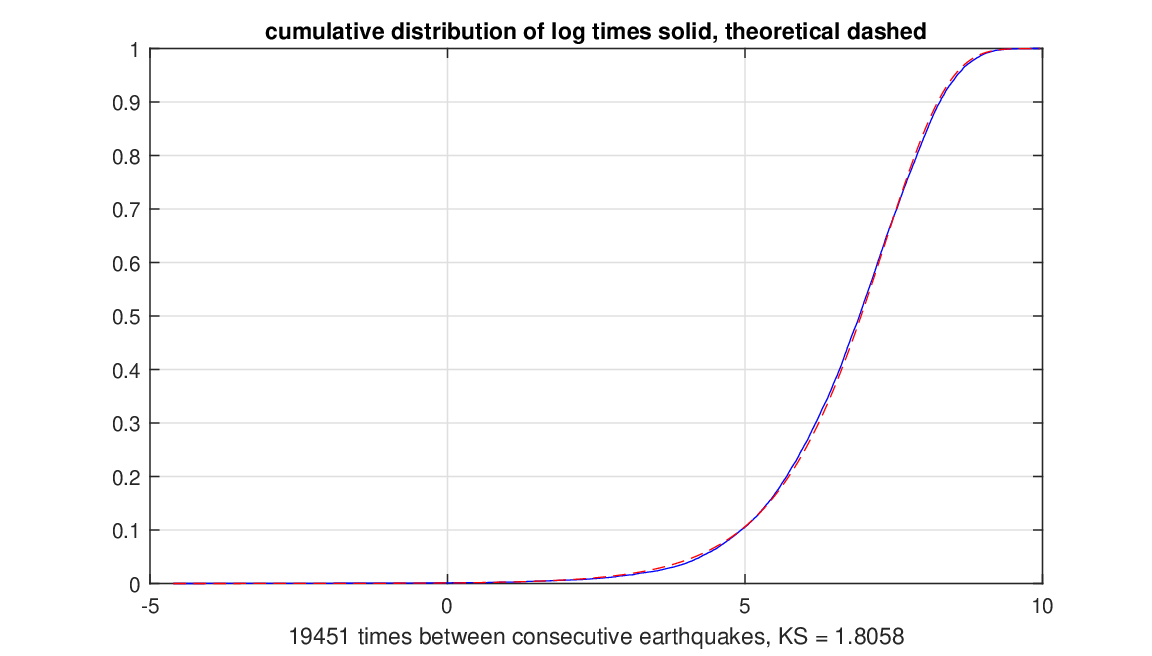}
    \caption{Earthquake Times Cumulative}
    \label{pop}
\end{figure}
\begin{lem}
If $X$ is log reflected Gumbel distributed then it has a Weibull 
density
\begin{equation}
q(x) = \frac{k}{\lambda} \left(\frac{x}{\lambda}\right)^{k-1} e^{-(x/\lambda)^k}
\end{equation}
where $\lambda = e^{-\mu}$ and $k = 1/\beta.$
\end{lem}
Proof: Equation (\ref{p}) gives $p(x) = q(e^x)e^x$ so the result follows from a direct computation.
\begin{remark}
Garavalia and colleagues \cite{garavaglia} have modelled earthquake interoccurence times from an Italian database spanning 20 years, using Weibull distrbutions. 
\end{remark}
\begin{lem}
The Weibel distribution is exponential iff $\beta = 1$ otherwise it equals the following mixture of exponential distributions where $f$ is a uniquely determined probability density:
\begin{equation}
q(x) = \int_0^\infty y f(y) e^{-yx} dy.
\end{equation}
Proof: Clearly $q(x) = -\frac{d}{dx} \mathcal L(f)(x)$ where $\mathcal L(f)$ is the Laplace transform of $f.$ Uniqueness follows since  
$\mathcal L(f)(0) = 1$ and the fact that the Laplace transform is invertible.
\end{lem}
\begin{remark}
Jewel \cite{jewell} has used mixture of exponential distributions to study  times of failure of units (patients, components) under observation. One can consider earthquakes as mechanical failures of geologic units in the Earth's crust. 
\end{remark}
\section{General Law for Random Variables}\label{sec5}
If $s$ consists of $n$ independent samples of an
$\mathbb R_+$-valued  random variable $X,$ then the law of large numbers (\cite{feller}, Chapter VII,  Lemma 1) implies that 
\begin{equation}\label{LLN}
\lim_{n \rightarrow \infty} Freq(R_{F,D}(s) = d) = Prob(R_{F,D}(X) = d),
\end{equation}
with probability $1$ and hence
\begin{equation}\label{Devdisc}
	\lim_{n \rightarrow \infty} E_{F,D}(s) = E_{F,D}(X) :=
\sum_{d=1}^D \left[ \, Prob(R_{F,D}(X) = d) - L_{F,D}(d) \, \right]^2.
\end{equation}
This crucial fact motivates our additional consideration regarding random variables and their probabilities, besides sampled data and their frequencies. Henceforth in this section we assume that $X$ is a continuously distributed $\mathbb R_+$-valued randon variable with density $q : \mathbb R_+ \mapsto [0,\infty).$ Then $\ln X$ is continuously distributed with density 
$p : \mathbb R \mapsto [0,\infty)$ given by
\begin{equation}\label{p}
p(x) = \lim_{\Delta x \rightarrow 0} \frac{Prob(\ln X \in [x,x+\Delta x))}{\Delta x}
= \lim_{\Delta x \rightarrow 0} \frac{Prob(X \in [e^x,e^x e^{\Delta x}))}{\Delta x} = 
q(e^x)e^x.
\end{equation}
Define the $\ln F$-periodized function
\begin{equation}\label{tildep}
	\widetilde p(x) := \sum_{j \in \mathbb Z} p(x + j\ln F).
\end{equation}
Clearly the restriction of $\widetilde p$ to the interval $[0,\ln F)$ is a probability density.  
Since
\begin{equation}\label{IFD2}
	R_{F,D}(x) = d \hbox{ iff } \ln x \in  I_{F,D}(d) + \mathbb Z \ln F,  \ \ 
			x \in \mathbb R_+; \, d = 1,...,D
\end{equation}
where 
\begin{equation}\label{IFD}
	I_{F,D}(d) := [\ln (1+(d-1)(F-1)/D), \ln (1+d(F-1)/D)), \ d = 1,...,D,
\end{equation}
it follows that
\begin{equation}\label{IFD3}
	Prob(R_{F,D}(X) = d) = \int_{I_{F,D}(d)} \widetilde p(x) \, dx.
\end{equation}
If $\widetilde p$ is constant then $\widetilde p = \frac{1}{\ln F},$ 
\begin{equation}\label{GL}
	Prob(R_{F,D}(X) = d) = L_{F,D}(d),
\end{equation}
and $E_{F,D}(X) = 0$ for all integers $D \geq 2.$
\begin{theo}\label{integral}
If $X$ is a $R_+$--valued continuously distributed random variable and $\ln X$ has density $p$ and $\widetilde p$ is defined by (\ref{tildep}) then
\begin{equation}\label{ineqE1}
DE_{F,D}(X) \leq (F-1)\int_0^{\ln F} e^{-x} \, \left[\widetilde p(x) - \frac{1}{\ln F}\right]^2\, dx.
\end{equation}
If $D_1$ divides $D_2$ then
\begin{equation}\label{ineqE2}
D_1\times E_{F,D_1}(X) \leq D_2\times E_{F,D_2}(X).
\end{equation}
If $\widetilde p$ is Riemann integrable, then
\begin{equation}\label{limE}
 \lim_{D \rightarrow \infty} D E_{F,D}(X) = 
	(F-1)\int_0^{\ln F} e^{-x} \, \left[\widetilde p(x) - \frac{1}{\ln F} \right]^2\, dx.
\end{equation}
\end{theo}
Proof: (\ref{ineqE1}) follows from the Schwarz inequality
$$\Bigg| \, Prob(R_{F,D}(X) = d) - L_{F,D}(d) \, \Bigg|^2 = 
\Bigg|\int_{I_{F,D}(d)} \left(\widetilde p(x) - \frac{1}{\ln F} \right) \, e^{-x/2}\, e^{x/2} \, dx  \, \Bigg|^2$$
$$\leq \int_{I_{F,D}(d)} e^xdx \, \int_{I_{F,D}(d)} \left(\widetilde p(x) - \frac{1}{\ln F} \right)^2 e^{-x}\, dx = \frac{F-1}{D} \int_{I_{F,D}(d)} \left(\widetilde p(x) - \frac{1}{\ln F} \right)^2 e^{-x}\, dx.$$
(\ref{ineqE2}) follows from Lemma (\ref{lem1}) and the law of large numbers (\ref{LLN}).
(\ref{limE}) follows from the definition of the Riemann integral and that fact that for large $D$ (\ref{Lbounds}) gives 
$\hbox{ length } I_{F,D}(d) = L_{F,D}(d) \approx \frac{F-1}{D + d(F-1)}, \ d = 1,...,D.$
\\ \\
The periodized normal densities, also called the wrapped normal distributions, are well known to be expressed by the Jacobi theta function. Figures 5-8 show the plots of the $\l F$-periodicised functions $\widetilde p$ for $F = 2, 8, 32, 128$ where $p$ is a normal density with mean $7.2170$ and standard deviation $1.831$ corresponding to our model for the US populations.The max-min values of $\widetilde p$ increase rapidly with increasing $F$ but are extremely small for $F = 2$ and $F = 8.$
\begin{figure}
    \centering
    \includegraphics[width=4in]{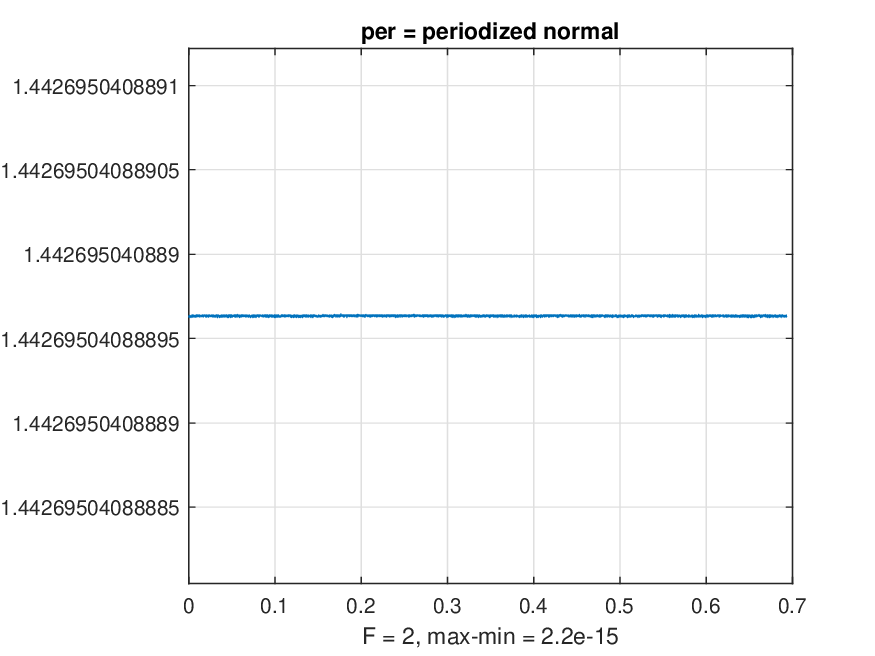}
    \caption{Periodized Normal Density}
    \label{pop}
\end{figure}
\begin{figure}
    \centering
    \includegraphics[width=4in]{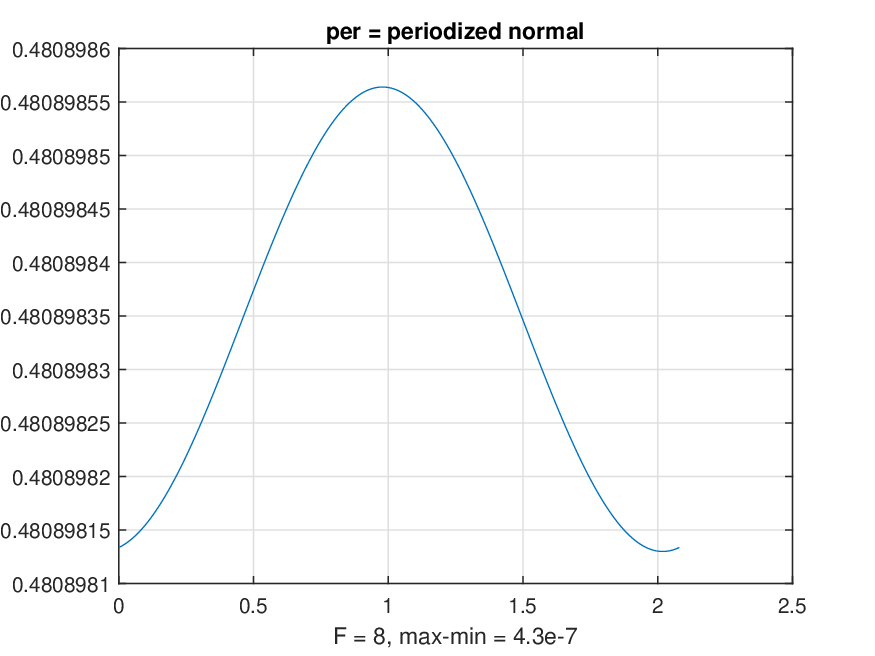}
    \caption{Periodized Normal Density}
    \label{pop}
\end{figure}
\begin{figure}
    \centering
    \includegraphics[width=4in]{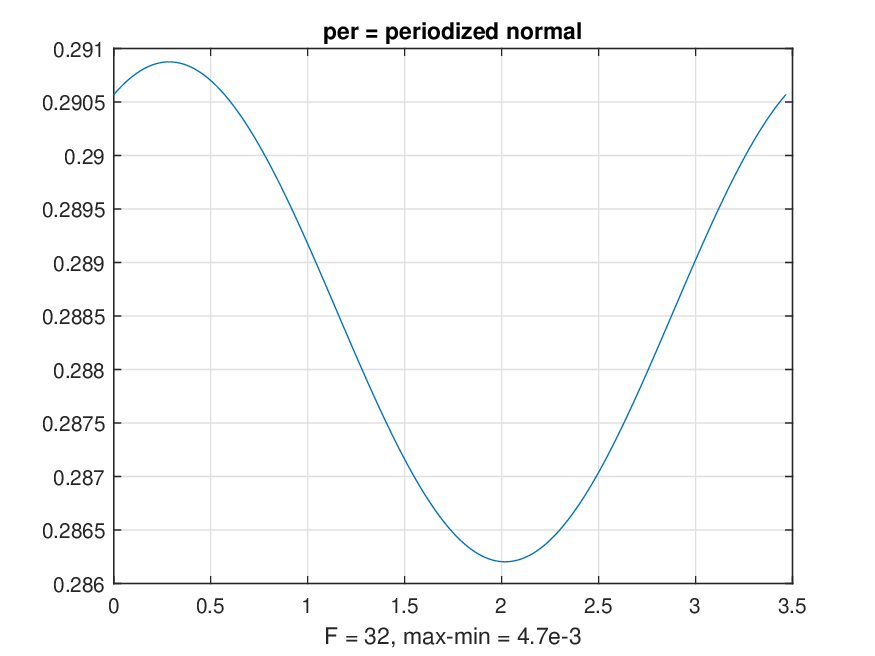}
    \caption{Periodized Normal Density}
    \label{pop}
\end{figure}
\begin{figure}
    \centering
    \includegraphics[width=4in]{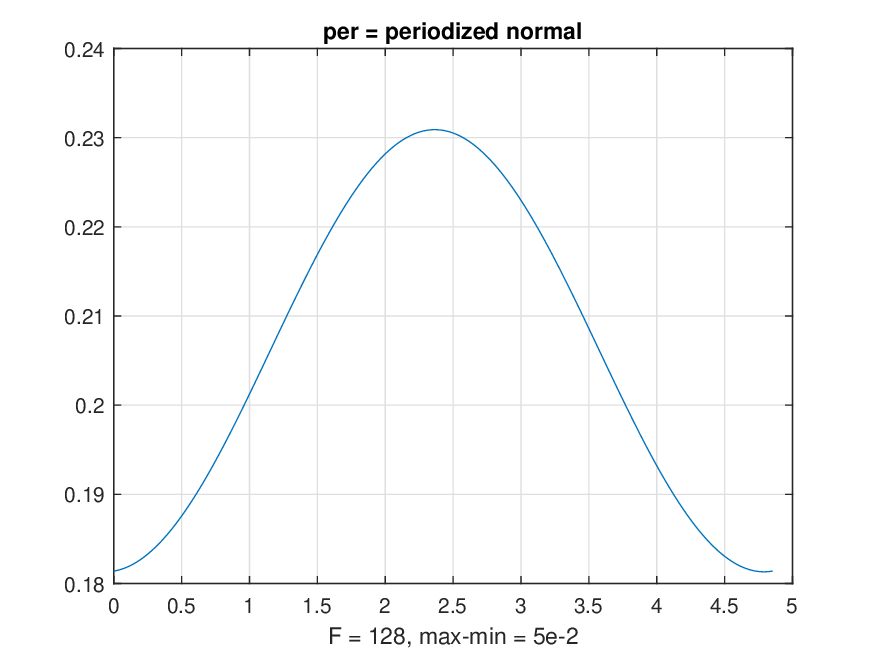}
    \caption{Periodized Normal Density}
    \label{pop}
\end{figure}
\begin{defi}\label{FourierTransform}
The Fourier transform of a density function $p : \mathbb R \mapsto [0,\infty)$ is the function $\widetilde p: \mathbb R \mapsto \mathbb C$ defined by
$\widehat p(y) := \int_{\mathbb R} p(x) \, e^{-2\pi i yx}\, dx.$
\end{defi}
\begin{lem}\label{FTdistributions}
The normal density
$$p(x) = \frac{1}{\sigma \sqrt {2\pi}} \exp(-\frac{1}{2\sigma^2}(x-\mu)^2)$$ 
has Fourier transform
$$\widehat p(y) = \exp(-2\pi i \mu y)\, \exp(-2\pi^2 \sigma^2 y)$$ 
which never vanishes. 
The reflected Gumbel density 
$$p(x) = \frac{1}{\beta} \exp (z - e^z), \ z = \frac{x + \mu}{\beta}$$
has Fourier transform
\begin{equation}\label{FTrefGumbeleqn}
\widehat p(y) = e^{2\pi i \mu y}\, \Gamma(1-2\pi i \beta y).
\end{equation}
which never vanishes.
\end{lem}
Proof: The Fourier transform of the normal density is computed by completing the square and then replacing the 
contour of integration $\mathbb R$ by a contour $\mathbb R + 2\pi i \sigma.$ For the reflected Gumbel density 
$\widehat p(y) := \int_{\mathbb R} p(x)\, e^{-2\pi i yx} \,  dx = e^{2\pi i \beta \mu y} \int_{\mathbb R} \exp (z - e^z) \, e^{-2\pi i \beta z y} \, dz$ \\
$= e^{2\pi i y \ln \mu} \int_0^\infty e^{-u} \, u^{(1-2\pi i y) - 1} du = 
e^{2\pi i y \ln \mu} \Gamma(1-2\pi i \beta y).$
The Gamma function never vanishes since it satisfies the well known identity
$$\Gamma(z)\, \Gamma(1=z) = \frac{\pi}{\sin \pi z}, \ \ z \in \mathbb C.$$
\begin{lem}\label{FourierInversion} Fourier Inversion Theorem: 
If $f :\mathbb R\rightarrow \mathbb C$ is integrable and sufficiently smooth 
and $h(x) := \int_{\mathbb R} f(x) \, e^{2\pi i yx}\, dy,$ 
then $\widehat h = f.$
\end{lem}
Proof: See (\cite{feller}, p. 509-510).
\begin{lem}\label{FTp1}
$\widetilde p = \frac{1}{\ln F}$ is constant iff the Fourier transform  satisfies
\begin{equation}\label{FTp2}
\widehat p(j/\ln F) = 0, \ \ j \in \mathbb Z \backslash \{0\}.
\end{equation}
\end{lem}
Proof: $\widetilde p$ has periodic $1/\ln F$ so has a Fourier series expansion 
$\widehat p(x) = \sum_{k\in \mathbb Z} c_k \, e^{2\pi i k x/\ln F}$ with
$c_k = \frac{1}{\ln F} \int_{[0,\ln F)} \widehat p(x) e^{-2\pi i k x/\ln F}.$ Substituting (\ref{tildep}) into this expression gives
$$c_k = \int_{[0,\ln F)} \sum_{j \in \mathbb Z} p(x+j \ln F) e^{-2\pi i k x/\ln F} \, dx$$  
$$= \int_{[0,\ln F)} \sum_{j \in \mathbb Z} p(x+j \ln F) e^{-2\pi i k (x+j\ln F)/\ln F} \, dx$$
$$= \sum_{j \in \mathbb Z} \int_{[0,\ln F)+j\ln F} p(x+j \ln F) e^{-2\pi i k (x+j\ln F)/\ln F} \, dx$$
$$ = \int_{\mathbb R} p(x) e^{-2\pi i k x/\ln F} \, dx= \widehat p(k/\ln F), \ \ k \in \mathbb Z.$$
\begin{defi}
A random variable $X$ is $F$-perfect if $E_{F,D}(X) = 0$ for every integer $D \geq 2.$
\end{defi}
The following is a slight generalization of "Benford’s law compliance theorem" derived by Steven Smith (\cite{smith}, p. 716).
\begin{lem}\label{perf}
For a continuously distributed random variable $X$ let $p$ be the density function of $\ln X$ and let $F > 1.$ Then $X$ is $F$-perfect iff $\widehat p(j/\ln F) = 0$ for every nonzero integer $j.$
\end{lem}
Proof: Follows directly from Equation (\ref{limE}) in Theorem \ref{integral} (\ref{limE}) and from Lemma \ref{FTp1}.
\begin{theo}\label{nonexistence1}
No random variable is $F_j$-perfect for every $F_j$ if $F_j \rightarrow \infty.$
For every $K > 0$ there exists a random variable $X$ that is $F$-perfect whenever 
$\ln F \leq 1/K.$ Lognormal and log reflected Gumbel distributed random variables are 
never $F$-perfect.
\end{theo}
Proof: If $X$ is $F_j$-perfect for every $j$ then 
$\widehat p(0) = \lim_{j \rightarrow \infty} \widehat p(1/\ln F_j) = 0$ but this contradicts the fact that $\widehat p(0) = \int_{\mathbb R} p(x)dx = 1.$  
To prove the second assertion define
$$
h(x) := \begin{cases}
\frac{1}{\sqrt K} \hbox{ if } x \in [-K/2,K/2] \\
0 \hbox{ otherwise}. 
\end{cases}
$$
and the convolution
$$
f(x) := \int_{\mathbb R}h(x-y)h(y)dy = 
\begin{cases}
1 + x/K \hbox{ if } x \in [-K,0] \\
1 - x/K \hbox{ if } x \in [0,K] \\
0 \hbox{ otherwise}.
\end{cases}
$$
Then define
$$
p(x) := \int_{\mathbb R} f(y) e^{2\pi i yx}\, dy = 
\bigg|\int_{\mathbb R} h(y) e^{2\pi i yx}\, dy\bigg|^2 
= \frac{1}{K} \left(\frac{\sin(\pi K x)}{\pi x}\right)^2.$$ 
Lemma \ref{FourierInversion} implies that
$\widehat p = f$ so $\widehat p(j/\ln F) = 0$ for 
$j \in \mathbb Z \backslash \{0\}$ and $\ln F \leq  1/K.$ Let $X$ be a random variable such that the density of $\ln X$ equals $p.$
The assertions about lognormal and log reflected Gumbel distributions follow from Lemmas \ref{FTdistributions} and \ref{perf}
\\ \\
The three tables below display values of $L_{F,5}(d), F = 2, 8, 32; d = 1,...,5$ and the 
Prob$(R_{F,D}(X) = d),  F = 2, 8, 32; d = 1,...,5$ for the lognormal and log reflected Gumbel distributed random variable $X$ whose parameters were chosen to equal those for US populations and times between successive earthquakes. These were the probability models discussed in Sections 4 and 5. The last columns contain the sum of squared errors.
$$
\begin{array}{ccccccc}
  L_{2,5}     & 0.26303441 & 0.22239242 & 0.19264508 &  0.16992500 & 0.15200309 & 0\\
  lognormal  & 0.26303441 & 0.22239242 & 0.19264508 &  0.16992500 & 0.15200309 & 1.1e-28\\
logGumbel   & 0.26303431 & 0.22239316 & 0.19264535 &  0.16992461 & 0.15200257& 1.1e-12
\end{array}
$$
$$
\begin{array}{ccccccc}
  L_{8,5}     & 0.42101147 & 0.22098834 & 0.15083740 & 0.11465147 & 0.09251133  & 0\\
  lognormal  & 0.42099290 & 0.22116695 & 0.15084018 & 0.11457438 & 0.09242559 &  4.6e-8\\
logGumbel   & 0.42311092 & 0.21018579 & 0.15025476 & 0.11896115 & 0.09748739 & 1.6e-4
\end{array}
$$
$$
\begin{array}{ccccccc}
  L_{32,5}   & 0.56959938 & 0.17923284 & 0.10972413 &  0.07930348 & 0.06214017 & 0\\
  lognormal  & 0.58122693 & 0.16012001 & 0.10727135 &  0.08330820 & 0.06807350 & 5.6e-4 \\
logGumbel   & 0.67229291 & 0.13437162 & 0.07473944 &  0.06235662 & 0.05623941 & 1.4e-2
\end{array}
$$
\begin{remark}
Comparing these tables with the tables for natural data sequences shows that probability
densities may better approximate the generalized law than the frequencies of natural data that they model.
This may be useful for density estimation that is a crucial tool in statistical pattern recognition, machine intelligence,
and data science.
\end{remark}
\section{Appendix 1: Kossovsky's Partition}\label{sec6}
In (\cite{koss2}, Section 7, Chapters 122 and 123)  Kossovsky partioned positive numbers  as
\begin{equation}\label{KP1}
\mathbb R_+ = (0,w) \cup \bigcup_{d = 1}^D \bigcup_{j = 0}^\infty   [1+(d-1)(F-1)/D,1+d(F-1)/D)wF^j
\end{equation}
where $w > 0$ is extremely small. Since 
$w = F^Lc$ where $c \in [1,F)$ and the integer $L << 0$ are unique, we get
\begin{equation}\label{KP2}
\mathbb R_+ = (0,F^Lc) \cup \bigcup_{d = 1}^D \bigcup_{j = L}^\infty [1+(d-1)(F-1)/D,1+d(F-1)/D)F^jc.
\end{equation}
Taking the limit as $w \rightarrow 0$ gives $L \rightarrow -\infty$ hence
\begin{equation}\label{KP3}
\mathbb R_+ = \bigcup_{d = 1}^D c\Sigma_{F,D}(d)
\end{equation}
where $\Sigma_{F,D}(d)$
 is defined in (\ref{SigmaFD}). So this corresponds to scaling the data by the factor $1/c$ which does not effect the Benford compliance of the data. However, for $F = D+1,$ and $x > 0,$ $x \in c\Sigma_{F,D}(d)$ iff the first digit of the base $F$ representation of $x/c$ equals $d.$
\\ \\
In (\cite{koss2}, Section 7, Chapters 124-129) Kossovsky used his partition (\ref{KP1}) to initially
define 
\begin{equation}\label{LFDlim}
	L_{F,D}(d) :=\lim_{L \rightarrow \infty} Prob(R_{F,D}(X_{w,L}) = d), \ d = 1,...,D
\end{equation}
where $X_{w,L}$ is the random variable with density $k/x$ on $[w,L]$ 
and $k = 1/\ln(L/w).$ He derived an expression for $Prob(R_{F,D}(X_{w,L}) = d), d = 1,...,D$ as the ratio  of two quantities both converging to $\infty.$ George Andrews proved that its limit equals the expression for $L_{F,D}(d)$ in (\ref{LFD}).
\section{Appendix 2: MATLAB Programs}\label{sec7}
function [GLvals,LFD,Diff,EFD] = GLnorm(F,D,mu,sigma) \\
Inputs and Outputs \\
F : base \\
D : number of bins \\
mu : mean of normal density \\
sigma : standard deviation of normal density \\
LFD : ideal GL values \\
GLvals : actual GL values \\
Start of Algorithm \\
r = 0:D; r = 1+r*(F-1)/D; r = log(r); \\
LFD = (r(2:D+1)-r(1:D))/log(F); \\
M = zeros(21,D+1); \\
LF = log(F); \\
for j = 1:101 \\
    M(j,1:(D+1)) = r + (j-51)*LF; \\
end \\
M = (M-mu)/sigma; \\
s = sign(M); \\
eM = 0.5+0.5*s.*erf(s.*M); \\
reM = sum(eM); \\
GLvals = reM(2:D+1)-reM(1:D); \\
Diff = GLvals-LFD; \\
EFD = Diff*Diff'; \\
end
\\ \\
function [GLvals,LFD,Diff,EFD] = GLGumbel(F,D,mu,beta) \\
Inputs and Outputs \\
F : base \\
D : number of bins \\
mu : mode (max argument) (= mean + beta*gamma) \\
beta : (= sigma*sqrt(6)//pi) \\
density(x) = (1/beta)*exp[w - exp(w)], w = (x-mu)/beta \\
LFD : ideal GL values \\
GLvals : actual GL values \\
Start of Algorithm \\
r = 0:D; r = 1+r*(F-1)/D; r = log(r); \\
LFD = (r(2:D+1)-r(1:D))/log(F); \\
M = zeros(21,D+1); \\
LF = log(F); \\
for j = 1:101 \\
    M(j,1:(D+1)) = r + (j-51)*LF; \\
end \\
M = (M-mu)/beta; \\
cM = 1-exp(-exp(M)); \\
rcM = sum(cM); \\
GLvals = rcM(2:D+1)-rcM(1:D); \\
Diff = GLvals-LFD; \\
EFD = Diff*Diff'; \\
end
\\ \\
function [x,per,mm] = periodizednormal(sigma,m,F) \\
period = log(F);N = 10000;dx = period/N; dy = period/(N-1);x = 0:dy:period; \\
a= -0.5/sigma$^2$; b = 1/(sigma*sqrt(2*pi)); J = size(x,2); \\
K1 = round(m-50*sigma); K2 = round(m+50*sigma); y = x-m; \\
for j = 1:J \\
    per(j) = 0; \\
    for k = K1:K2 \\
        d = b*exp(a*(y(j)-k*period)$^2$); \\
        per(j) = per(j)+d; \\
    end \\
end \\
plot(x,per) \\
title('per = periodized normal') \\
grid \\
mm = max(per)-min(per); \\
end
\\ \\
{\bf Acknowledgment} We thank the distinguished mathematician George Andrews 
for his ingenious proof which enabled the entire discussion 
of this article. Andrews is well known for his discovery of Ramanujan's lost notebook at Trinity College's library, Cambridge in 1976, and for his extensive work on Ramanujan's research. Andrews stated recently that his computation of the expression for $L_{F,D}$ in the GLORQ was inspired by this work. Accordingly both authors indirectly owe gratitude to Ramanujan of India as well.

\end{document}